\documentclass[aps,onecolumn,preprintnumbers,nofootinbib,notitlepage]{revtex4}
\usepackage{amssymb, amsthm}
\usepackage{amsmath}
\usepackage{graphicx}
\usepackage{caption}
\usepackage{subcaption}
\usepackage{hyperref}
\usepackage{dcolumn}
\usepackage{bm}%
\usepackage{xcolor}
\usepackage{float}
\usepackage{hyperref}%
\hypersetup{
    colorlinks,
    citecolor=orange,
    filecolor=black,
    linkcolor=blue,
    urlcolor=blue
}
\usepackage{doi}

\begin{document}
\title{Dark matter for Majorana neutrinos in a $\mathbb{Z}_4$ symmetry}

\author{Leon M.G. de la Vega,$^{1,2}$\footnote{Contact author: leonmgarcia@phys.ncts.ntu.edu.tw} Patrick J. Fitzpatrick,\footnote{Contact author: patrickf@campus.technion.ac.il} Rolando Martinez-Ramirez,$^{3,2}$\footnote{Contact author: romartne@uni-mainz.de} and Eduardo Peinado$^{2,4}$\footnote{Contact author: epeinado@fisica.unam.mx}}
\affiliation{%
$^1$Physics Division, National Center for Theoretical Sciences, National Taiwan University, Taipei 106319, Taiwan.\\
$^2$Instituto de F\'{\i}sica, Universidad Nacional Aut\'onoma de M\'exico, A.P. 20-364, Ciudad de M\'exico 01000, Mexico.\\
$^3${\rm PRISMA}$^+$ Cluster of Excellence and 
Institute for Nuclear Physics, Johannes Gutenberg-University, 55099 Mainz, Germany.\\
$^4$Departamento de F\'isica, Centro de Investigaci\'on y de Estudios Avanzados del Instituto Polit\'ecnico Nacional, 07000 Ciudad de México, Mexico.}%

\begin{abstract}
In the minimal Scotogenic model with fermionic dark matter, the neutrino Yukawa matrix is fine-tuned to satisfy current bounds on $\mu \rightarrow e \gamma$ while producing dark matter through thermal freeze-out. One way to avoid this bound is to open a new annihilation channel for dark matter (DM) with a scalar field that breaks the lepton number spontaneously. This introduces a physical Goldstone boson $a$, which can constitute a light relic and opens up the lepton flavor violating channel $\mu\rightarrow e a$. In this work, we introduce a model that contains the new annihilation channel for fermionic dark matter but avoids the introduction of the Goldstone boson by using a discrete $\mathbb{Z}_4$ instead of a continuous symmetry. In this way, cosmological constraints on the light relic and experimental limits on lepton flavor violating decays to the Goldstone boson are avoided. We analyze the experimental viability of the model, identifying the region of parameter space where lepton flavor violating constraints are not saturated, DM has the correct relic density, and DM direct detection experiments are sensitive to the DM candidate.

\end{abstract}

\maketitle

\section{Introduction}

The Standard Model (SM) accurately describes the fundamental particles and their interactions via the strong and electroweak forces. Nevertheless, massless neutrinos are considered in its formulation, contradicting neutrino oscillation experiments that establish the existence of at least two massive neutrinos~\cite{PhysRevLett.20.1205,Super-Kamiokande:1998kpq,SNO:2002tuh,KamLAND:2002uet,K2K:2006yov,MINOS:2006foh,KamLAND:2008dgz,DoubleChooz:2011ymz,DayaBay:2012fng,RENO:2012mkc,T2K:2011ypd,T2K:2014ghj}. The mechanism that provides masses to neutrinos remains an open question within the SM framework.

Meanwhile, observations of large-scale structure formation and the shape of the cosmic microwave background (CMB) anisotropy power spectrum conclusively establish the existence of dark matter (DM) in the Universe, providing a value for its relic density given by~\cite{Planck:2018vyg}
\begin{equation}
    \Omega_c h^2 = 0.120 \pm 0.0012 \text{ at } 68\%\ \text{ confidence level} .
\label{eq:PlanckKing}
\end{equation}
The SM does not include a viable DM candidate.

It is possible to address both major theoretical shortcomings in SM extensions where the neutrino masses are generated at the radiative level, and an extra symmetry both forbids tree-level neutrino mass terms and provides a DM candidate (see, e.g., \cite{Krauss:2002px,Cheung:2004xm}). In this class of SM extensions, the smallness of neutrino masses is addressed naturally due to extra suppression from loop factors. Additionally, the lightest particle charged under the extra symmetry becomes stable and constitutes the DM.

The scotogenic model~\cite{Ma:2006km} is a paradigmatic example of this class of SM extensions, where a discrete $\mathbb{Z}_{2}$ symmetry is imposed, and neutrino masses are generated by one-loop processes. In this case, the DM candidate is the lightest particle charged under the $\mathbb{Z}_2$ symmetry. While an additional $\mathbb{Z}_2$ symmetry composes the scotogenic model, distinct $\mathbb{Z}_n$ symmetries can be used to stabilize the dark sector in a similar framework, leading to different phenomenology (see, e.g.~\cite{Srivastava:2019xhh,Yaguna:2024jor}). For example, a residual $\mathbb{Z}_{2n}$ symmetry can arise from the breaking of a continuous symmetry already present in the SM, such as $U(1)_{B-L} $\cite{Bonilla:2018ynb, CentellesChulia:2019gic}, giving DM and neutrino phenomenology similar to the minimal scotogenic case.

\begin{figure}
    \centering
    \includegraphics[width=0.5\linewidth]{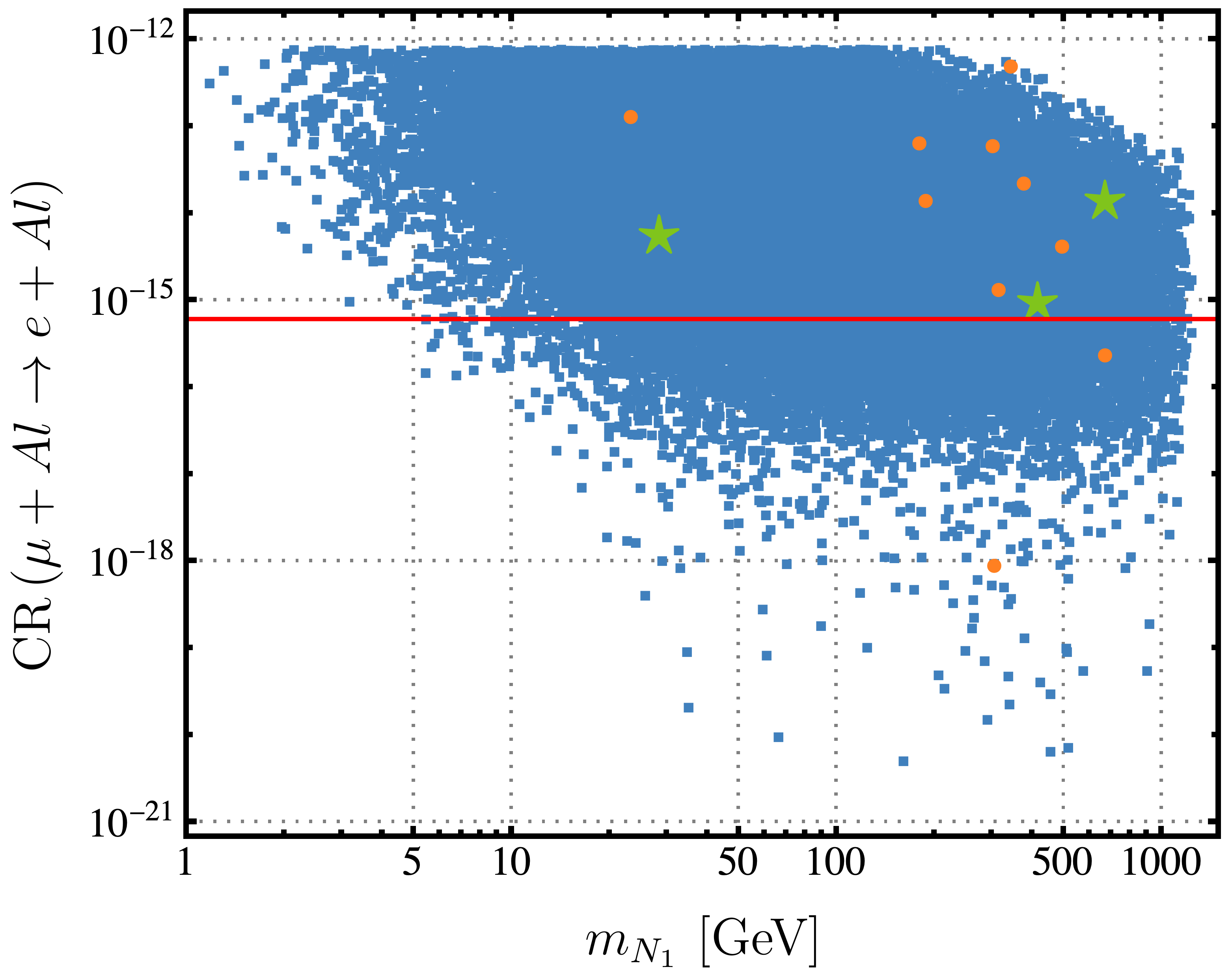}
    \caption{Results of a scan of the minimal scotogenic model as defined in \cite{Ma:2006km}. Fixing the neutrino oscillation parameters to their best fit point in the normal hierarchy \cite{Esteban:2020cvm}, we performed a search for points with real, perturbative Yukawa couplings, $|Y_p|<3$, fermion dark matter mass between 1.0 GeV and 1.25 TeV, and compliant with current LFV limits from $l_i\rightarrow l_j \gamma$ \cite{MEG:2016leq,BaBar:2009hkt,Belle:2021ysv}, $l_i\rightarrow 3 l_j $ \cite{SINDRUM:1987nra,Hayasaka:2010np}, and $\mu N \rightarrow e N $ \cite{SINDRUMII:2006dvw} processes. We show this sample of benchmark points in the two-dimensional space of dark matter mass and $\mu$ to $e$ conversion rate (CR) in aluminum. In this plot, blue squares correspond to the full sample of LFV compliant points regardless of its prediction of DM relic density, orange circles correspond to benchmark points that do not predict DM overabundance and green stars correspond to points with a calculated DM relic density in the range $\Omega_N h^2=0.1196\pm5\%$. The red line shows the projected sensitivity of run 1 of the Mu2e experiment at Fermilab \cite{Mu2e:2022ggl} to $\text{CR }(\mu+ Al\rightarrow e + Al)$.}
    \label{fig:scotoscan}
\end{figure}

When the DM candidate in the minimal scotogenic model is the lightest dark Majorana fermion, the annihilation channel that controls DM production requires either sizable Yukawa couplings to SM leptons or a coannihilation enhancement through the $\eta N_1\rightarrow SM SM$ or $N_1 N_j\rightarrow SM SM$ processes. For the first case, the large Yukawa couplings produce observably large rates for lepton flavor violating (LFV) processes such as $\mu \rightarrow e \gamma$, unless the Yukawa couplings are fine-tuned so that the $\tau$ coupling of DM is much larger than the electron and muon couplings. Although current constraints on the parameter space without coannihilation enhancement do not rule out the model completely, future sensitivity of LFV probes will disfavor this scenario \cite{Vicente:2014wga,Ahriche:2017iar,Karan:2023adm}.
To put this into perspective, we have performed a numerical scan of the scotogenic model with DM mass between 1 GeV and 1 TeV using the public codes SARAH (versions 4.15.2 and 4.11.0)~\cite{Staub:2013tta}, SPheno (4.0.2) \cite{Porod:2003um,Porod:2014xia} and micrOmegas (5.3.41)~\cite{Belanger:2018ccd}. We find that from a random sample of 42433 benchmark points (BP) satisfying all current LFV experimental bounds, only $0.03\%$ (13) do not predict DM overabundance. We observe that the future LFV search with the most potential to explore the remaining parameter space is Mu2e at Fermilab, which will probe the $\mu$ to $e$ conversion rate in aluminum down to $6.2 \times 10^{-16}$ \cite{Mu2e:2022ggl}, see Fig. \ref{fig:scotoscan}. From this cursory look at the scotogenic model, without considering coannihilations or fine-tuning of the Yukawa matrix phases, we can see that the parameter space of the model is already tightly constrained when considering both dark matter relic density and current and future bounds on LFV observables. When the spectrum of the minimal scotogenic model contains a dark scalar or dark fermion with a mass very close to the fermion DM mass, the DM annihilation cross section is enhanced by the new channels, which helps the DM observables to decouple from LFV predictions \cite{Karan:2023adm}.
{\color{red}Another} viable solution is to introduce a new scalar field that couples to the dark fermions, dynamically generates their mass, and opens up a new DM annihilation channel \cite{Bonilla:2019ipe}. If the DM mass, in this case, is generated by the spontaneous breaking of a global symmetry, the Majoron is also generated, which gives rise to new phenomenology constrained by LFV, Higgs-to-invisible decays, and possible cosmological constraints on the new light relics \cite{DeRomeri:2022cem}.

In this work, we present a new possibility to explain the origin of neutrino masses and the production of DM: it is possible to dynamically generate light fermion masses by breaking a discrete $\mathbb{Z}_4$ symmetry, which prevents the generation of neutrino masses at tree level. Through the spontaneous breaking of the $\mathbb{Z}_4$ symmetry to a $\mathbb{Z}_2$ symmetry, the DM is stabilized, and neutrino masses are generated at one-loop order, as in the scotogenic model. For a type I seesaw implementation of this idea, see~\cite{Bhattacharya:2018ljs}. The lightest particle, which is odd under the $\mathbb{Z}_2$ subgroup, provides a DM candidate with either bosonic or fermionic nature. Meanwhile, this scenario avoids the prohibitive LFV constraints on DM production in the minimal scotogenic case, as well as the phenomenological consequences of the Majoron associated with the dynamical generation of DM masses by the breaking of a continuous global symmetry, as proposed in previous studies.

This model provides a uniquely economical way to generate the observed neutrino masses as well as the DM relic abundance, in a way that is experimentally viable against stringent constraints on an extended lepton sector. Furthermore, the experimentally viable window can provide sharp targets, which will be explored in near-future direct DM searches at the electroweak scale.

This paper is organized as follows. In Sec.~\ref{sec:Model} we introduce the model by enumerating the symmetries and field content. We write down the interaction terms and calculate the physical spectrum. We identify theoretical and experimental constraints on the scalar spectrum. With this, in Sec.~\ref{sec:ExperimentalViability} we enumerate experimental constraints on the model, namely, DM observables and charged LFV limits. We define the region of the parameter space over which we will perform a numerical scan. In Sec.~\ref{sec:NumericalResults} we present the results of the scan, identifying the region where the correct DM relic density is obtained while satisfying other experimental limits. Finally, in Sec.~\ref{sec:Conclusions} we conclude on the success of the model in presenting a viable DM candidate and generating neutrino masses and mixings without saturating limits on LFV rates or the abundance of light relics.

\section{The model}
\label{sec:Model}
We propose a SM extension that includes three right-handed sterile neutrinos $N_k$, a scalar $\eta$, transforming as $(2,1/2)$ under the $SU(2)_L\times U(1)_Y$ SM gauge group, and two scalars $\sigma$ and $\phi$, which are singlets under the SM gauge group. These new particles transform nontrivially under an additional $\mathbb{Z}_4$ symmetry, with the $\mathbb{Z}_4$ charges listed in Table~\ref{Particle_content}.

\begin{table}[h]
\begin{center}
\begin{tabular}{| c || c | c | c |}
  \hline 
&   \hspace{0.1cm}  Fields     \hspace{0.1cm}       &    $SU(2)_L \otimes U(1)_Y$            &  
   \hspace{0.4cm} $\mathbb{Z}_{4}$ \hspace{0.4cm}   \\
\hline \hline

Fermions  &  &  &	  \\	
&
 $L_i$        	  &    ($2, {-1/2}$)       	  &	 {$1$}  \\
&   $e_{R_i}$       &   ($1, {-1}$)    &  	 {$1$}\\
&   $N_k$       &   ($1, {0}$)         &  	 {$i$}\\
 & & & \\
\hline \hline

Scalars   & & & \\
&
 $H$  		 &  ($2, {1/2}$)        & {$1$}    \\
& $\eta$          	 &  ($2, {1/2}$)        &  $i$    \\		
& $\sigma$          	 &  ($1, {0}$)          &  {$i$}       \\
& $\phi$             &  ($1, {0}$)           &	{$-1$} \\
& & & \\
    \hline
  \end{tabular}
\end{center}

\caption[Particle content in the two-loop model]{Particle content of the model with $i\in\{e, \mu, \tau\}$ and $k \in \{1,2,3\}$. All the fields listed are $SU(3)_C$ singlets. Fields with imaginary charges under $\mathbb{Z}_4$ become odd under $\mathbb{Z}_{2}$ after SSB, and the lightest among them provides a stable DM candidate. The scalar field $\phi$ is chosen to be real.}
 \label{Particle_content}
\end{table}

The most general scalar potential of this model is given by 

\begin{equation} \label{scalar_potential_oneloop}
\begin{aligned}
    V = &\mu^2_H \left( H^{\dagger} H \right) + \mu^2_{\eta} \left( \eta^{\dagger} \eta \right) + \mu^2_{\sigma} \left( \sigma^* \sigma \right) + \mu^2_{\phi} \left( \phi^2 \right) \\
    &+ \lambda_1 \left( H^{\dagger} H \right)^2 + \lambda_2 \left( H^{\dagger} H \right) \left( \eta^{\dagger} \eta \right) + \lambda_3 \left( H^{\dagger} \eta \right) \left( \eta^{\dagger} H \right) + \lambda_4 \left( H^{\dagger} H \right) \left( \sigma^{*} \sigma \right) + \lambda_5 \left( H^{\dagger} H \right) \phi^2 \\
    &+ \lambda_6 \left( \eta^{\dagger} \eta \right)^2 + \lambda_7 \left( \eta^{\dagger} \eta \right) \left( \sigma^{*} \sigma \right) + \lambda_8 \left( \eta^{\dagger} \eta \right) \phi^2 + \lambda_9 \left( \sigma^{*} \sigma \right)^2 + \lambda_{10} \left( \sigma^{*} \sigma \right) \phi^2 \\
    &+ \lambda_{11} \phi^4 + \bigg[ \xi \left( H^{\dagger} \eta \right) \sigma \, \phi + \kappa_1 \left( H^{\dagger} \eta \right) \sigma^* + \kappa_2 \, \sigma^2 \phi + \rm h.c. \bigg],
\end{aligned}
\end{equation}
where
\begin{equation}
    H = \begin{pmatrix}
        H^+ \\
        H^0
    \end{pmatrix} \quad \text{and} \quad \eta= \begin{pmatrix}
        \eta^+ \\
        \eta^0
    \end{pmatrix}
\end{equation}
represent the SM Higgs and the scalar inert doublet, respectively.

The Yukawa interactions, relevant for neutrino masses, compatible with the SM and $\mathbb{Z}_{4}$ symmetries are
\begin{equation} \label{yukawa_oneloop}
    \mathcal{L}_{Y} = Y_{p}  \overline{L} \,  \tilde{\eta}  N + Y_v  \overline{N^c} N \phi + \rm H.c.
\end{equation}
The coupling coefficients $\xi, \kappa_1$, $\kappa_2$, $\lambda_a \, (a = 1,...,11)$, $Y_{p}$ and $Y_{v}$ in Eqs.~\ref{scalar_potential_oneloop} and~\ref{yukawa_oneloop} are assumed to be real.

\begin{figure}[!]
    \centering
    \includegraphics[width=0.57\linewidth]{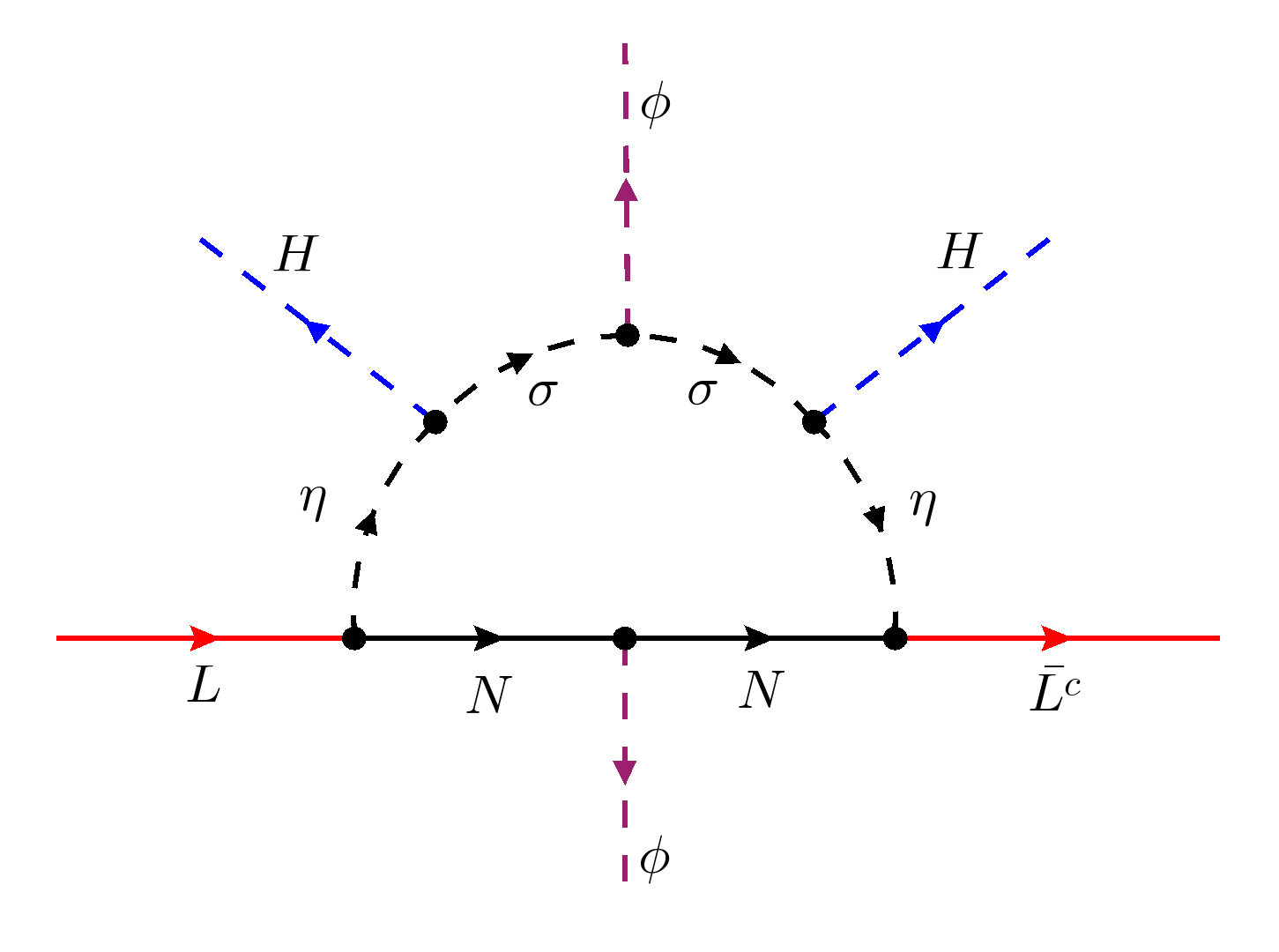}
    \caption{One-loop Feynman diagram for neutrino mass generation. After SSB, both $\phi$ and $H$ acquire nonzero vacuum expectation values and Majorana neutrino masses are generated. The particles inside the loop constitute the dark sector.}
    \label{One_loop_diagram}
\end{figure}

    In the stable vacuum, both $\phi$ and $H$ acquire nonzero vacuum expectation values, and the $\mathbb{Z}_{4}$ and electroweak symmetries are spontaneously broken. Majorana masses are then generated at leading order through one-loop processes, as shown in Fig.~\ref{One_loop_diagram}.

\subsection{Mass spectrum}
\label{sec:MassSpectrum}
We parametrize the neutral scalar fields $H$, $\eta$, $\sigma$, and $\phi$ around the vacuum in terms of their $CP$-even and $CP$-odd components as
\begin{equation} \label{scalar_oneloop}
\begin{aligned}
     H^0 &\rightarrow \dfrac{1}{\sqrt{2}} \left( v_H + h^0_R + i \, h^0_I \right), \\
     \eta^0 &\rightarrow \dfrac{1}{\sqrt{2}} \left( v_\eta + \eta^0_R + i \, \eta^0_I \right), \\
     \sigma &\rightarrow \dfrac{1}{\sqrt{2}} \left( v_\sigma + \sigma_R + i \, \sigma_I \right), \\
     \phi &\rightarrow \dfrac{1}{\sqrt{2}} \left( v_\phi + \phi_R \right),
\end{aligned}
\end{equation}
where $v_H$ is the SM Higgs vacuum expectation value. The scalar potential minimum corresponds to
\begin{equation}
    H^{0} \rightarrow \frac{1}{\sqrt{2}} v_{H} \quad \text{and} \quad \phi \rightarrow \frac{1}{\sqrt{2}} v_{\phi},   
\end{equation}
which spontaneously breaks the symmetry $SU(2)_L\times U(1)_Y\times \mathbb{Z}_4\rightarrow U(1)_{QED}\times \mathbb{Z}_2$. The stable vacuum configuration is given by
\begin{equation} \label{tadpoleeqs_oneloop}
\begin{aligned}
v_\eta&=v_\sigma=0,\\
    \mu^2_H &= - v_H^2 \, \lambda_1 - \dfrac{1}{2} v_\phi^2 \, \lambda_5, \\
    \mu^2_\phi &= - v_\phi^2 \, \lambda_{11} - \dfrac{1}{2} v_H^2 \, \lambda_5.
\end{aligned}
\end{equation}

Spontaneous symmetry breaking (SSB) induces mass mixing between the scalars in the visible sector ($h^{0}_{R}$ and $\phi_{R}$) through the Higgs-portal interaction $\lambda_{5}$. Mass mixing is also separately induced between the scalar dark sector ($\eta^0_R$ and $\sigma_R$) and the pseudoscalar dark sector ($\eta^0_I$ and $\sigma_I$). Meanwhile, the fields with imaginary charge $i$ under $\mathbb{Z}_4$ are odd under the residual $\mathbb{Z}_{2}$ symmetry after SSB, with the lightest particle being a stable DM candidate.

The mass matrix for the visible sector is given by
\begin{equation}
     \mathcal{M}_{\text{Visible}} = \begin{pmatrix}
        2 \, v_H^2 \, \lambda_1 && v_H \, v_\phi \, \lambda_5 \\
        v_H \, v_\phi \, \lambda_5 && 2 \, v_\phi^2 \, \lambda_{11}
    \end{pmatrix},
\end{equation}
which is diagonalized by the $2 \times 2$ rotation matrix
\begin{equation}
    \mathcal{O}_{\text{Visible}} = \begin{pmatrix}
        \cos{\theta_V} & \sin{\theta_V} \\
        -\sin{\theta_V} & \cos{\theta_V}
    \end{pmatrix},
\end{equation}
where
\begin{equation}
    \tan{\left( 2 \, \theta_V \right)} = \dfrac{v_H \, v_\phi \, \lambda_5}{  v_\phi^2 \, \lambda_{11} - v_H^2 \, \lambda_1}.
\end{equation}
Thus, the rotation between the interaction and mass eigenstates $h$ and $\varphi$ is given by
\begin{equation} \label{visible_mix}
    \begin{pmatrix}
        h \\
        \varphi
    \end{pmatrix} = \mathcal{O}_{\text{Visible}} \begin{pmatrix}
        h^0_R \\
        \phi_R
    \end{pmatrix},
\end{equation}
with mass eigenvalues
\begin{align}
    m_{h}^2 &= v_H^2 \, \lambda_1 + v_\phi^2 \, \lambda_{11} - \sqrt{( v_H^2 \, \lambda_1 - v_\phi^2 \, \lambda_{11})^2 + (v_H \, v_\phi \, \lambda_5)^2 }, \label{mh_mass_eq} \\
    m_{\varphi}^2 &= v_H^2 \, \lambda_1 + v_\phi^2 \, \lambda_{11} + \sqrt{( v_H^2 \, \lambda_1 - v_\phi^2 \, \lambda_{11})^2 + (v_H \, v_\phi \, \lambda_5)^2 }.  \label{mvarphi_mass_eq}
\end{align}
We identify one of these mass eigenstates $h$ as the physical Higgs field, with a mass of $m_{h} = 125.25 \text{ GeV}$.

The dark sector mass matrix is given by
\begin{equation}
     \mathcal{M}_{DM_X} = \begin{pmatrix}
        a && b_X \\
        b_X && c_X
    \end{pmatrix},
\end{equation}
where
\begin{equation}
\begin{aligned}
    a &= \dfrac{1}{2} \bigg [ v_H^2 \, ( \lambda_2 + \lambda_3 ) + v_{\phi}^2 \, \lambda_8 + 2 \, \mu_{\eta}^2 \bigg ], \\
    b_X &= \dfrac{1}{2} v_H \left( \sqrt{2} \, \kappa_1 + \zeta_X \, v_{\phi} \, \xi \right), \\
    c_X &= \dfrac{1}{2} \bigg [ \zeta_X \, 2\sqrt{2} \, v_{\phi} \, \kappa_2 + v_{\phi}^2 \, \lambda_{10} + v_H^2 \, \lambda_4 + 2 \, \mu^2_{\sigma} \bigg ].
\end{aligned}
\end{equation}
Here $X = \{ R,I \} $ denotes scalar $(R)$ and pseudoscalar $(I)$ dark sector fields, with $\zeta_{R} = 1$ and $\zeta_I = -1$. Similar to the visible case, the mass matrices are diagonalized by a $2 \times 2$ rotation matrix $\mathcal{O}_{DM_X}$ with
\begin{equation}
    \tan{\left( 2 \, \theta_X \right)} = \dfrac{2 \, b_X}{c_X-a}.
\end{equation}
The rotation between the interaction and mass eigenstates $X_{1}$ and $X_{2}$ is 
\begin{equation} \label{dark_mix}
    \begin{pmatrix}
        X_{1} \\
        X_{2}
    \end{pmatrix} = \mathcal{O}_{DM_X} \begin{pmatrix}
        \eta^0_X \\
        \sigma_X
    \end{pmatrix},
\end{equation}
with the corresponding masses
\begin{align}
    m_{X_{1}}^2 &= \dfrac{1}{2} \left( a + c_X - \sqrt{(a - c_X)^2 + 4 \, b_X^2 } \right), \\
    m_{X_{2}}^2 &= \dfrac{1}{2} \left( a + c_X + \sqrt{(a - c_X)^2 + 4 \, b_X^2 } \right).
\end{align}

The mass for the charged scalar particle $\eta^\pm$ is given by
\begin{equation}
    m_{\eta^\pm}^2 = \dfrac{1}{2} \left( v_H^2 \, \lambda_2 + v_{\phi}^2 \, \lambda_8 \right) + \mu_{\eta}^2.
\end{equation}

SSB also generates the masses of fermions $N_k$ through the Yukawa interaction
\begin{equation} \label{eq_mN}
    m_{N_k} = \sqrt{2} \, (Y_v)_{kk} \, v_\phi.
\end{equation}
We work in the basis where fermions $N_k$ do not mix with each other, implying a diagonal Yukawa matrix $Y_v$. 

\subsection{Neutrino masses}
\label{sec:NeutrinoMasses}
SSB generates light neutrino masses through the one-loop mechanism shown in Fig.~\ref{One_loop_diagram}. The corresponding $3 \times 3$ neutrino mass matrix can be written as

\begin{widetext}
\begin{equation}
    \left( \mathcal{M}_\nu \right)_{\alpha \beta} = \sum_i^3 \dfrac{(Y_p)_{\alpha i} \, (Y_p)_{i \beta}}{32 \pi^2} M_{N_i} \sum_{A = 1}^4 V_A (\theta_X) V_A (\theta_X) \left[ \frac{m_A^2 \log(m_A^2/M_{N_i}^2)}{m_A^2-M_{N_i}^2} \right],
\end{equation}\end{widetext}
where 

\begin{equation}
\begin{aligned}
    m_1 &= m_{R_{1}}, \quad \quad m_2 = m_{R_{2}}, \\
    m_3 &= m_{I_{1}}, \quad \quad m_4 = m_{I_{2}};
\end{aligned}
\end{equation}
and
\
\begin{equation} \label{eq:DarkMixCoeff}
\begin{aligned}
    V_1 &= \cos \left( \theta_R \right), \quad \quad   V_2 = - \sin \left( \theta_R \right), \\
    V_3 &= - i \cos \left( \theta_I \right), \quad \quad  V_4 = i \sin \left( \theta_I \right).
\end{aligned}
\end{equation}

\subsection{Dark matter production}
\label{sec:DarkMatterProduction}
After SSB the Majorana fermions $N_{k}$, as well as the neutral scalar and pseudoscalar dark sector fields $X_{1}$ and $X_{2}$ are odd under the remaining $\mathbb{Z}_{2}$ symmetry. Therefore, the lightest of these states is stable, providing a DM candidate. Here we will consider the phenomenology of the case in which the DM is the lightest Majorana fermion, $N_{1}$.

\begin{figure}[!]
    \centering
    \includegraphics[width=0.65\linewidth]{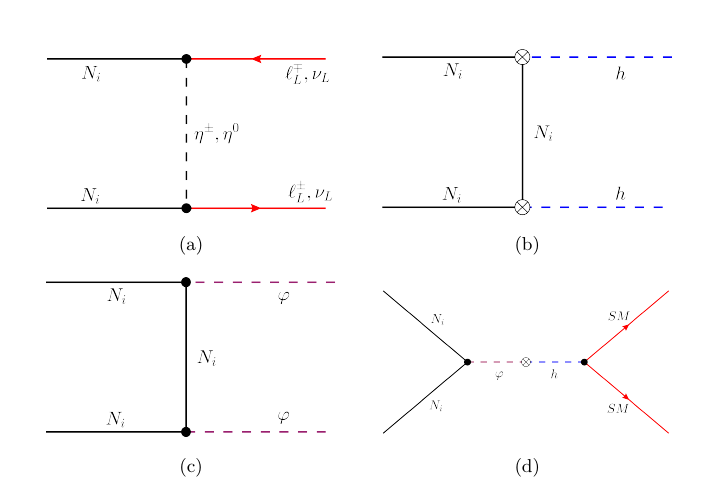}
    \caption{Tree-level DM annihilation channels including a) $t$-channel annihilations to SM leptons, b) $t$-channel annihilations to Higgs bosons via Higgs-portal mixing, c) $t$-channel annihilations to scalars $\varphi$, and d) $s$-channel annihilations to SM via Higgs-portal mixing.}
    \label{fig:DMAnnihilations}
\end{figure}

We consider thermally produced DM with a mass in the GeV$-$TeV scale, where a wealth of current and near-future direct detection experiments are particularly sensitive. We will characterize the experimental targets of DM production at the sub-GeV scale in a forthcoming study. At the electroweak scale, DM annihilates primarily via the channels shown in Fig.~\ref{fig:DMAnnihilations}.

DM that is thermally produced by $t$-channel annihilations to SM leptons through their Yukawa interaction with the scalar dark sector, as shown in Fig.~\ref{fig:DMAnnihilations}(a), encounters prohibitively strong experimental constraints from LFV processes \cite{Bonilla:2019ipe,Toma:2013zsa,Vicente_2015}. Meanwhile, weakly interacting massive particle (WIMP) DM produced via $t$-channel annihilations to SM particles through the Higgs-portal mixing $\lambda_{5}$, as shown in Fig.~\ref{fig:DMAnnihilations}(b), predicts a large visible mixing angle $\sin \theta_{V}$, which is excluded by experimental constraints from direct detection \cite{LZ:2022lsv} and Higgs-to-invisible decays~\cite{ATLAS:2020kdi}. The same is true for WIMP DM thermally produced via $s$-channel annihilations, shown in Fig.~\ref{fig:DMAnnihilations}(d), except in the experimentally viable resonance region where either $2 m_{N_{1}} \sim m_{h}$ or $2 m_{N_{1}} \sim m_{\varphi}$. In the remaining window, DM is thermally produced through $2 \rightarrow 2$ annihilations to scalars, $NN \rightarrow \varphi \varphi$, shown in Fig.~\ref{fig:DMAnnihilations}(c), with scalars $\varphi$ subsequently decaying to the SM.

DM production through $NN \rightarrow \varphi \varphi$ annihilations exhibits two distinct parameter regions, the kinematically allowed region where $m_{N} > m_{\varphi}$, which controls DM production in the majority of the experimentally viable window, and the region where $m_{N} < m_{\varphi}$ and kinematically suppressed $NN \rightarrow \varphi \varphi$ annihilations, proceeding at finite temperature, require exponentially enhanced couplings $Y_{v} \sin \theta_{V}$ to thermally produce the DM relic density, relative to the kinematically allowed case. Kinematically suppressed $NN \rightarrow \varphi \varphi$ annihilations produce the DM relic abundance in a thin mass window $m_{\varphi}/m_{N} \gtrsim 1$ consistent with perturbative Yukawa couplings $Y_{v}$.

The results for DM production presented here are obtained by numerical integration of the Boltzmann equations over the entire experimentally viable parameter region, using micrOmegas (5.3.41)~\cite{Belanger:2018ccd}, as described in Sec.~\ref{sec:NumericalResults}.

\section{Experimental Viability}
\label{sec:ExperimentalViability}

\subsection{Boundedness conditions}
\label{sec:BoundednessConditions}

Perturbativity of the theory requires the scalar and Yukawa couplings of Eqs.~\ref{scalar_potential_oneloop} and~\ref{yukawa_oneloop} to satisfy
\begin{equation}
\begin{aligned}
    &|\lambda_a|, |\xi|, |Y_v^{kk}|^2, |Y_p^{ij}|^2 < 4 \, \pi, \\
    \quad &\text{with} \quad a = (1,...,11), \, k = (1,...,3) \, \text{and} \, \, i,j = (1,...,3). 
\end{aligned}
\end{equation}
The condition for the scalar potential to be bounded from below imposes the following additional constraints in the parameter spectrum~\cite{Branco:2011iw,Lindner:2016kqk}:
\begin{equation} \label{bounded_from_below}
\begin{aligned}
    &\lambda_{1,6,9,11} > 0, \quad \quad \lambda_2 \geq -2\, \sqrt{\lambda_1 \, \lambda_6}, \quad \quad \lambda_4 \geq -2\, \sqrt{\lambda_1 \, \lambda_9}, \quad \quad \lambda_5 \geq -2\, \sqrt{\lambda_1 \, \lambda_{11}}, \\
    &\lambda_7 \geq -2\, \sqrt{\lambda_6 \, \lambda_9}, \quad \quad \lambda_8 \geq -2\, \sqrt{\lambda_6 \, \lambda_{11}}, \quad \quad \lambda_{10} \geq -2\, \sqrt{\lambda_9 \, \lambda_{11}}, \quad \quad \lambda_2 + \lambda_3 \geq -2 \sqrt{\lambda_1 \lambda_6}.
\end{aligned}
\end{equation}

\subsection{Neutrino oscillation data}
\label{sec:NeutrinoOscillations}

The light neutrino mass matrix can be rewritten as
\begin{equation}
    (\mathcal{M}_{\nu})_{\alpha \beta} = \left( Y_p^T \, \Lambda \, Y_p \right)_{\alpha \beta},
\end{equation}
where the matrix $\Lambda = \text{diag}\left( \Lambda_{i} \right)$, with
\begin{equation}
\begin{aligned}
    \Lambda_i &=  \, \dfrac{ 1 }{32 \pi^2}  M_{N_i} \sum_{A = 1}^4 V_A (\theta_X) V_A (\theta_X) \left[ \dfrac{m_A^2 \log m_A^2 - M_{N_i}^2 \log M_{N_i}^2}{m_A^2 - M_{N_i}^2} \right] ,
\end{aligned}
\end{equation}
and the dark sector mixing parameters $V_{A} \left( \theta_{X} \right)$ are defined as in Eq.~\ref{eq:DarkMixCoeff}.

The neutrino mass matrix can be diagonalized as
\begin{equation} \label{diag_mass_majorana}
    U_{\rm PMNS}^T \, \mathcal{M}_{\nu} \, U_{\rm PMNS} = \hat{m}_{\nu} \equiv \begin{pmatrix}
        m_1 & 0 & 0 \\
        0 & m_2 & 0 \\
        0 & 0 & m_3
    \end{pmatrix} ,
\end{equation}
where $U_{\rm PMNS}$ is the Pontecorvo-Maki-Nakagawa-Sakata (PMNS) matrix. In the PDG parametrization, we can write the PMNS matrix in terms of three mixing angles and one complex phase,
\begin{equation}
    U_{\text{PMNS}} = \begin{pmatrix}
        c_{12}c_{13} & s_{12}c_{13} & s_{13}e^{-i \delta_{CP}} \\
        -s_{12} c_{23} - c_{12}s_{13} s_{23}e^{i \delta_{CP}} & c_{12} c_{23} - s_{12}s_{13} s_{23}e^{i \delta_{CP}} &c_{13}s_{23}\\
         s_{12} s_{23} - c_{12}s_{13} c_{23}e^{i \delta_{CP}} & -c_{12} s_{23} - s_{12}s_{13} c_{23}e^{i \delta_{CP}} &c_{13}c_{23}\\
    \end{pmatrix},
\end{equation}
where we have used the shorthand $c_{ij}=\cos\theta_{ij}$, $s_{ij}=\sin\theta_{ij}$.
Using the Casas-Ibarra parametrization \cite{Casas:2001sr}, the Yukawa matrix $\left( Y_p \right)_{i \alpha}$ can be written as
\begin{equation} \label{casas_ibarra_oneloop}
    Y_p = \sqrt{\Lambda}^{-1} R \, \sqrt{\hat{m}_{\nu}} \, U_{\rm PMNS}^{\dagger},
\end{equation}
where $R$ is a complex orthogonal matrix satisfying $R^T R = 1$. Using this parametrization, we are able to use the neutrino oscillation data, the scalar spectrum of the theory, and the matrix $R$ as input to determine the Yukawa matrix values. Neutrino oscillation experiments have measured quite precisely the mixing in the lepton sector, with several global fits agreeing on the allowed range of mixing angles \cite{deSalas:2020pgw,Capozzi:2021fjo,Esteban:2020cvm}. The remaining major unknowns in the three neutrino oscillation paradigm are the value of the $CP$ violating phase $\delta_{CP}$ and the sign of the atmospheric neutrino mass squared difference $\Delta m_{i3}^2$. 
The choice of $R$ does not affect the prediction for mixing angles and neutrino masses, but it does affect the prediction for the rates of LFV processes.
In this work, we aim to decouple the DM annihilation cross section in the early Universe from LFV constraints. In order to achieve this, our strategy for fitting oscillation data is to set constraints on the scalar spectrum so that the Yukawa couplings obtained in Eq.~\ref{casas_ibarra_oneloop} lead to small LFV rates without resorting to a fine-tuning of the Yukawa matrix structure through the choice of $R$. For this purpose, we take into account the cosmological limit on the sum of neutrino masses \cite{eBOSS:2020yzd}
\begin{equation}
    \sum_im_{\nu_i}< 0.13 \text{ eV} \quad \quad \text{CMB[TT,TE,EE+lowE] + BAO}.
\end{equation}
obtained by an analysis of CMB temperature (TT), temperature-polarization cross-correlation (TE), E-mode polarization (EE) anisotropy spectra, low power E-mode polarization (low E) and Baryon Acoustic Oscillation data (BAO).

\subsection{Electroweak precision constraints}
\label{sec:ElectroweakPrecisionConstraints}

The enriched scalar sector induces additional contributions to the electroweak oblique parameters at the one-loop level with respect to the SM. We calculate these one-loop contributions to \textit{S} and \textit{T} following \cite{Grimus:2008nb}. These parameters are tightly constrained by global fits of electroweak precision data; we consider the ranges \cite{PDG2024:2024pdg}
\begin{eqnarray}
    S&=&-0.05 \pm 0.07\\
    T&=&0.00\pm 0.06
\end{eqnarray}
with a positive 93\% correlation.

\subsection{Lepton flavor violating processes}
\label{sec:leptonflavorviolation}

The Yukawa interactions, which contribute to the generation of neutrino masses in this model, also produce LFV radiative decays through charged $\eta^{\pm}$-$N$ loops, as in the minimal scotogenic model \cite{Toma:2013zsa}. This leads to experimental constraints on LFV radiative decays $l_i\rightarrow  l_j \gamma$, three-body decays $l_i\rightarrow 3 l_j $, and $\mu$-$e$ conversion processes, $\mu+N\rightarrow e + N$, measured in scattering experiments.
We consider the strongest bound $l_i\rightarrow l_j \gamma$ to set the requirements for the scalar spectrum that leads to small $Y_p$ Yukawa couplings and, therefore, to a DM phenomenology dominated by the $Y_v$ couplings.\\
The $l_i\rightarrow l_j \gamma$ branching ratio is given by 
\begin{equation}
    BR(l_i\rightarrow l_j \gamma) =\frac{3(4\pi)^3\alpha_{em}}{4 G_F^2} |A_D|^2 BR(l_i\rightarrow l_j \nu_i\overline{\nu}_j),
\end{equation}
where $l_i\rightarrow l_j \nu_i\overline{\nu}_j$ is the SM, $W$ mediated charged lepton decay process, and 
\begin{equation}
    A_D=\sum_{a=1}^{3} (Y_p)_{ai}^*(Y_p)_{aj}\frac{F_2(m^2_{N_i}/m^2_{\eta})}{2(4\pi)^2 m^2_{\eta}} \; ,\; F_2(x)= \frac{1-6x+3x^2 +2x^3 -6x^2 \ln{x}}{6(1 - x)^4}.
\end{equation}
The current most sensitive experimental limit is 
\begin{equation}
     BR(\mu\rightarrow e \gamma) <3.1 \times 10^{-13},
\end{equation}
obtained from MEG-II data in combination with MEG data \cite{MEGII:2023ltw}.

\subsection{Higgs decay}
\label{sec:HiggsDecay}

The Yukawa interaction between DM and visible scalars in our model predicts an additional invisible decay channel for the SM Higgs boson, $h \rightarrow N N$, through the Higgs-portal coupling $\lambda_{5}$. Direct searches for invisible decay channels of the Higgs boson by ATLAS and CMS (e.g., \cite{CMS:2022qva, ATLAS:2020kdi,Biekotter:2022ckj}) therefore put direct experimental limits on our model when $m_{h} > 2 m_{N_{i}}$.

The $h \rightarrow N N$ decay rate is given by
\begin{equation}
    \Gamma_{h \rightarrow NN} = \frac{1}{8 \pi} m_{h} Y_{v}^{2} \sin^{2} \theta_{V} Re \left( 1 - \frac{4 m_{N_{1}}^{2}}{m_{h}^{2}} \right)^{3/2} .
\end{equation}

The current strongest constraints on the branching ratio $\text{Br}\left( h \rightarrow NN \right)$, obtained from the combined vector boson fusion and top mediated production of Higgs bosons at ATLAS \cite{ATLAS:2020kdi}, give $\text{Br}\left( h \rightarrow NN \right) < 0.11$ at 95$\%$ confidence level. This constraint directly translates to an upper bound on the product of Yukawa coupling and visible mixing angle, $Y_{v} \sin \theta_{V}$ for a given $m_{N_{1}}$. We project these experimental limits on the physical parameter region consistent with DM production and neutrino mass generation in Figs.~\ref{P30}$-$\ref{P500} and find that they are subdominant to direct detection constraints at the GeV$-$TeV scale.

\subsection{Direct detection}
\label{sec:DirectDetection}
The visible mixing of the SM Higgs and the singlet $\phi$ via the Higgs-portal coupling induces a tree-level contribution to the DM-nucleon spin-independent elastic scattering cross section $\sigma_{\textbf{SI}}$. The cross section is given by \cite{Matsumoto:2018acr}
\begin{equation}
    \sigma_{\textbf{SI}}= Y_v^2 \sin{\theta_V}^2\cos{\theta_V}^2\frac{m_{N}^2 m_{\text{nuc}}^4 f_N^2 }{\pi v_H^2(m_N+m_{\text{nuc}})^2}\left(m_\varphi^{-2}-m_{h}^{-2}\right)^2,
\end{equation}
where $f_N = f_{Tu} + f_{Td} + f_{Ts} + (2/9)f_{TG}$ with $f_{Tu} = 0.0153$, $f_{Td} = 0.0191$, $f_{Ts}= 0.0447$, and $f_{TG} = 0.921$ and $m_{\text{nuc}}$ is the mass of the nucleon. This sets a limit on the scalar mixing angle as a function of the DM mass, the scalar mass, and the Yukawa coupling $Y_v$. Roughly, assuming that the mixing angle is small, to avoid the recent LUX-ZEPLIN (LZ) constraint $\sigma\leq 10^{-47} \text{cm}^2$ \cite{LZ:2022lsv}, we find $Y_\nu\sin\theta_V<10^{-4}$ for $m_\phi=30$ GeV and $Y_\nu\sin\theta_V<10^{-3}$ for $m_\phi=100$, $500$ GeV. The recent LZ results give the strongest experimental constraints to the physical parameter region consistent with DM and neutrino masses, as shown in Figs.~\ref{P30}$-$\ref{P500}. 

\section{Numerical Results}
\label{sec:NumericalResults}
We perform a numerical analysis using the computational tools SARAH (versions 4.15.2 and 4.11.0)~\cite{Staub:2013tta}, SPheno (4.0.2) \cite{Porod:2003um,Porod:2014xia} and micrOmegas (5.3.41)~\cite{Belanger:2018ccd}. The model is first implemented in SARAH, which provides analytical expressions for the spectrum and couplings of the model and generates a source code for SPheno, where the analytical expressions are evaluated numerically given specific values of the Lagrangian parameters.
We use micrOmegas to solve the Boltzmann equations for DM production and calculate direct detection cross sections based on the numerical outputs obtained from SPheno.\footnote{For a pedagogical tutorial on these computational tools, see~\cite{Vicente:2015zba}.}

With the Majorana fermion $N_1$ constituting the DM, we perform a random scan over all physical parameters of the model, over the ranges displayed in Table~\ref{input_scan_dm}. 
\begin{table}[H]
\begin{center}
\begin{tabular}{| c |}
  \hline 
   \hspace{0.1cm}  Parameter   \hspace{0.1cm}         \\
\hline \hline
\\
$ \lambda_{2,3,4,6,7,8,9,10}  \in  [10^{-6}, 1]$ \\	
 \\
 $ (Y_v)_{11}   \in   [0.05, 1] $  \\	
  \\
$ \kappa_{1}   \in   [10^{-6}, 1]  \, \, \rm GeV$  \\	
  \\
$ \mu^2_\eta, \mu^2_\sigma    \in    [10^{7}, 10^{8}] \, \, \rm GeV^2 $ \\	
  \\
  $\kappa_{2} \in [10^{-2}, 1] \, \, \rm GeV$ \\ \\
  $\xi = 10^{- 13}$ \, \, \\ \\
$ m_{N_1}   \in   [10, 2000] \, \, \rm GeV $  \\	
  \\
    \hline
  \end{tabular}
\end{center}
\caption[Input parameters for the relic density scan]{Parameter ranges of the numerical scan.}
\label{input_scan_dm}
\end{table}
We consider the three cases of fixed $m_\varphi = 30$, $m_\varphi = 100$, and $m_\varphi = 500 \, \rm GeV$, which gives a representative characterization of DM production in each distinct thermal freeze-out regime over the entire electroweak scale. Over each numerical scan, the subset of all parameters, which is consistent with boundedness conditions, perturbativity, neutrino oscillation data, electroweak precision constraints, and LFV constraints, as described in Sec.~\ref{sec:ExperimentalViability}, is identified. For example, the Yukawa couplings $(Y_p)_{ij}$ are fixed to reproduce observed neutrino oscillation data and to be allowed by LFV constraints. In order to identify the experimentally viable window for DM production against direct detection constraints, for each distinct value of $m_{\varphi}$ we consider a different range for the Higgs-portal mixing $\lambda_5$,
\begin{align*}
   \lambda_5 &\in [10^{-5},10^{-3}] \text{ for } m_\varphi = 30, \\  
   \lambda_5 &\in [10^{-4},10^{-2}] \text{ for } m_\varphi = 100, \quad \text{and} \\   
   \lambda_5 &\in [10^{-3},10^{-1}] \text{ for } m_\varphi = 500 \, \rm GeV.
\end{align*}

We have set $\xi=10^{-13}$ to fix $\theta_R \simeq \theta_I$ for simplicity and without loss of generality. Furthermore, the small values of $\kappa_1$ lead to $\theta_{R/I}\ll 1$. In this limit, neutrino masses are determined dominantly by the $R_1$ and $I_1$ loops. We set a minimum mass difference between these states of $|m_{R_1}-m_{I_1}|> 0.1 \, \rm GeV$, which in Eq.~\eqref{casas_ibarra_oneloop} leads to $Y_p<10^{-2}$ for all entries of the matrix without resorting to fine-tuning through the choice of $R$. This mass difference is controlled with the parameter $\kappa_2$. With this, we obtain $BR(\mu\rightarrow e\gamma)<10^{-15}$, 2 orders of magnitude below the current limit set by MEG-II and 1 order of magnitude below its planned sensitivity. Crucially, the Yukawa couplings in this limit can be pushed even lower by tuning the value of $\kappa_2$ to increase the mass splitting $|m_{R_1}-m_{I_1}|$ without affecting DM phenomenology. The only observable sensitive to this splitting, besides the neutrino mass matrix, is the oblique parameter $T$.

\begin{figure}[H]
    \centering
    \includegraphics[width=0.8\linewidth]{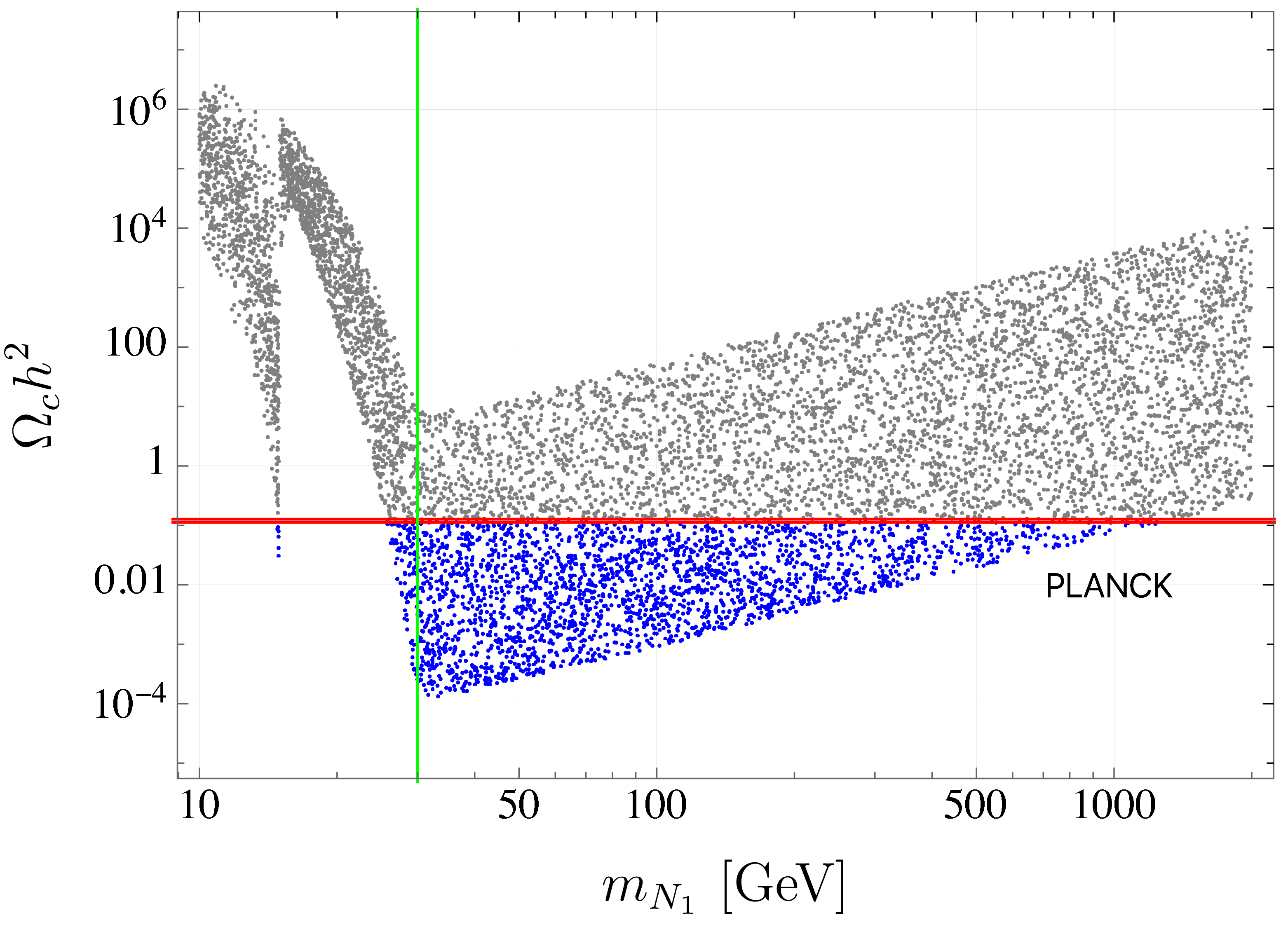}
    \caption{Relic density as function of $m_{N_1}$ for $m_\phi = 30 \rm \, GeV$. The red horizontal line represents the value of the relic density observed by Planck within a $3\sigma$ range. Blue points represent solutions with underabundance of DM relic density, gray points lead to overabundance. The green vertical line indicates where $m_{N_{1}} = m_{\varphi}$.}
    \label{rd_30}
\end{figure}

\begin{figure}[H]
    \centering
    \includegraphics[width=0.8\linewidth]{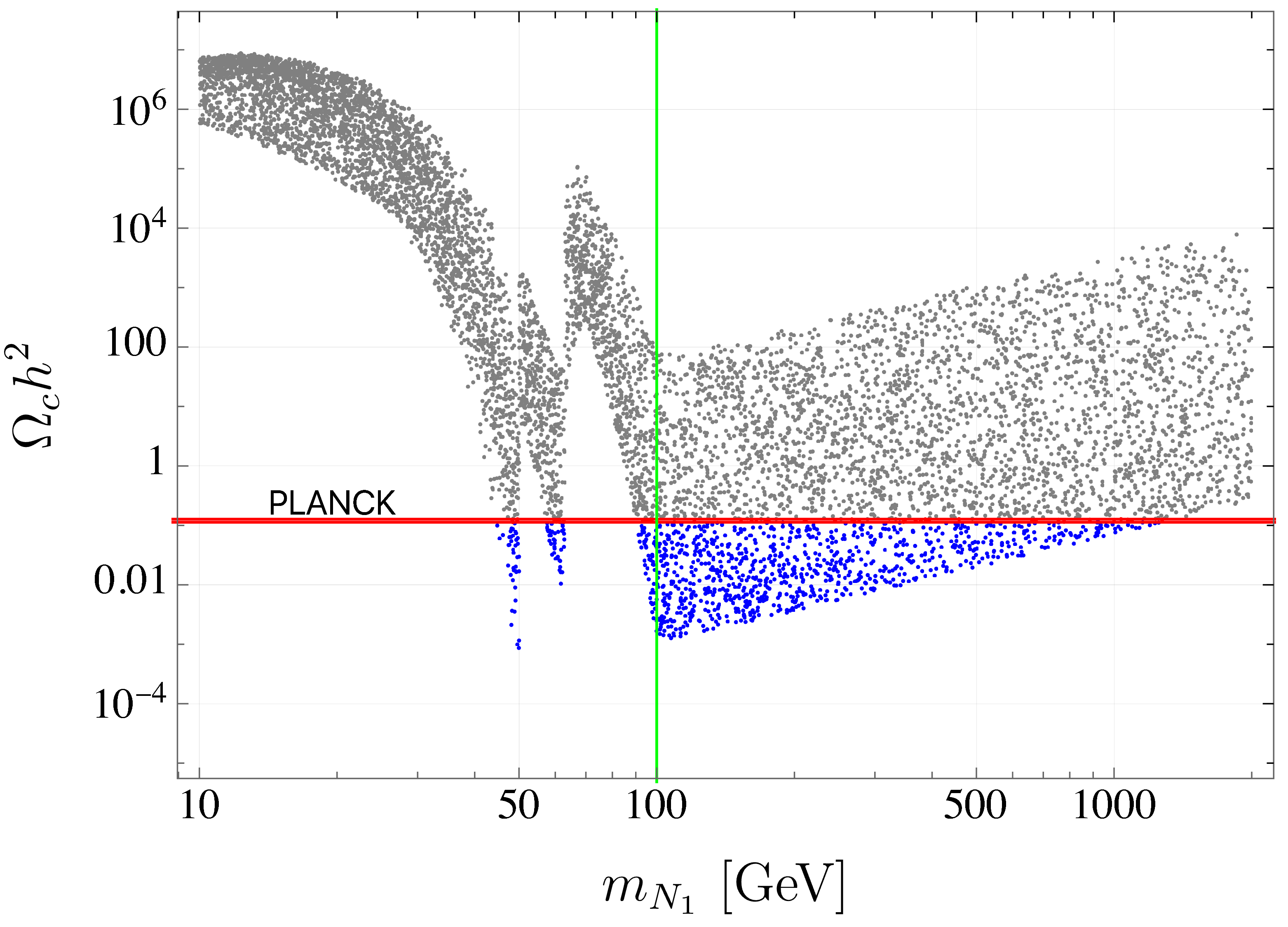}
    \caption{Relic density as function of $m_{N_1}$ for $m_\phi = 100 \rm \, GeV$. Same color code as Fig.~\ref{rd_30}.}
    \label{rd_100}
\end{figure}

\begin{figure}[H]
    \centering
    \includegraphics[width=0.8\linewidth]{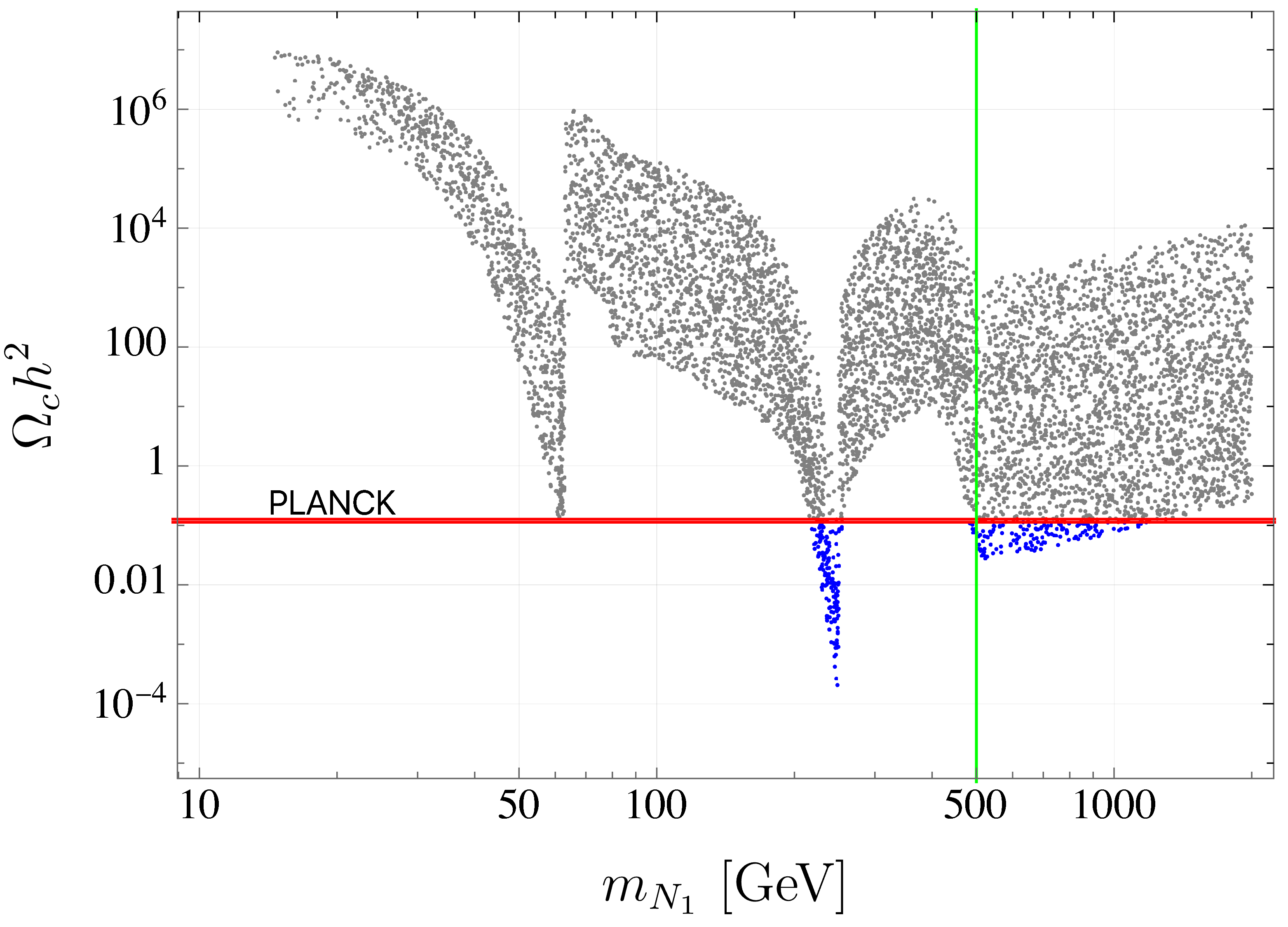}
    \caption{Relic density as function of $m_{N_1}$ for $m_\phi = 500 \rm \, GeV$. Same color code as Fig.~\ref{rd_30}.}
    \label{rd_500}
\end{figure}

Results for DM production over data consistent with the experimental limits described above are shown in Figs.~\ref{rd_30}$-$\ref{rd_500}. 
The value of the DM relic abundance predicted by experimentally viable data is shown as a function of DM mass. Horizontal red lines represent values of the relic density within a $3\sigma$ range of the result observed by Planck~\cite{Planck:2018vyg}, identified in Eq.~\ref{eq:PlanckKing}. Gray points represent data that are inconsistent with DM production, predicting DM overabundance, while blue points represent data that are consistent with the production of at least a fraction of the DM relic abundance.

Most of the data in Figs.~\ref{rd_30}$-$\ref{rd_500} produce the DM abundance through $t$-channel annihilations to scalars $\varphi$ [shown in Fig.~\ref{fig:DMAnnihilations}(c)], which then promptly decay to SM particles. In the region where $m_{N_{1}} < m_{\varphi}$, DM is produced through kinematically suppressed $NN \rightarrow \varphi \varphi$ annihilations in only a thin parameter region where $\frac{m_{N_{1}}}{m_{\varphi}} \sim 1$. At lower values of $\frac{m_{N_{1}}}{m_{\varphi}} < 1$ the kinematic suppression of $NN \rightarrow \varphi \varphi$ annihilations leads to an overabundance of DM. There are also thin regions at $2m_{N_{1}} \sim m_{h}$ and $2m_{N_{1}} \sim m_{\varphi}$ where the DM relic abundance is produced by $s$-channel DM annhilations [shown in Fig.~\ref{fig:DMAnnihilations}(d)] through an on-shell resonance.

\begin{figure}[H]
        \centering
        \includegraphics[height=4in]{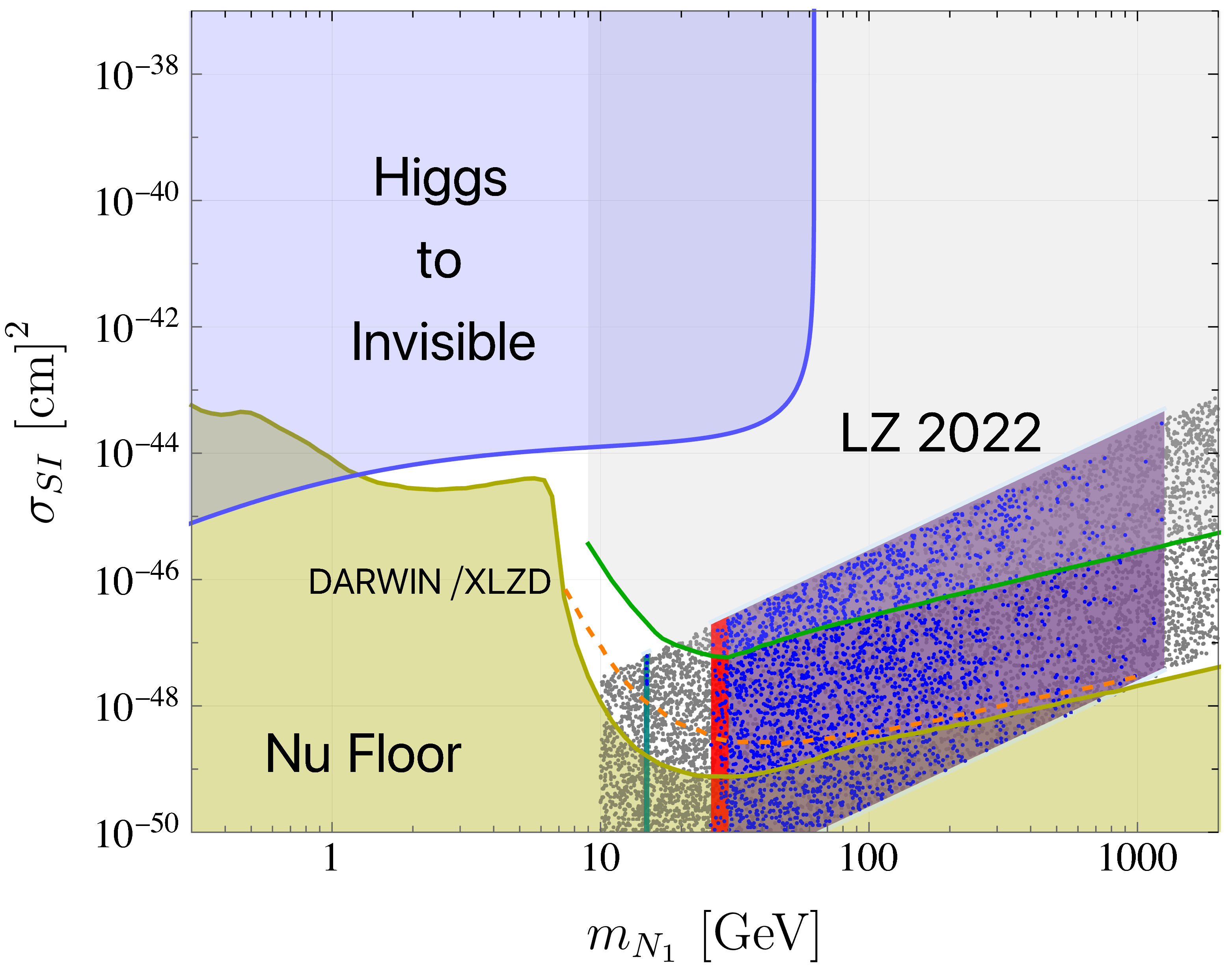}
        \caption{Direct detection in the model for $m_\varphi = 30 \, \rm GeV$. All data from Fig.~\ref{rd_30} are shown. Colored bands correspond to each regime of DM production. Purple band indicates where the relic abundance of DM is produced by kinematically allowed $NN \rightarrow \varphi \varphi$ annihilations, red band indicates where the relic abundance is produced through kinematically suppressed $NN \rightarrow \varphi \varphi$, and turquoise bands correspond to production of the DM abundance by $s$-channel DM-to-SM annihilations through an on-shell $h$ or $\varphi$ resonance. It is worth noting that the upper line of the bands corresponds to $\lambda_5=10^{-3}$, and the lower line is for $\lambda_5=10^{-5}$; if we continue lowering the values for $\lambda_5$, we will find regions with lower cross section. The green line represents the LZ 2022 constraint on spin independent direct detection \cite{LZ:2022lsv}, the purple line indicates the limit on Higgs-to-invisible decays \cite{ATLAS:2020kdi}, the orange dashed line represents the DARWIN projection constraint~\cite{Baudis:2024jnk}, the yellow region is the neutrino floor for xenon.} 
        \label{P30}
    \end{figure}

\begin{figure}[H]
        \centering
        \includegraphics[height=4in]{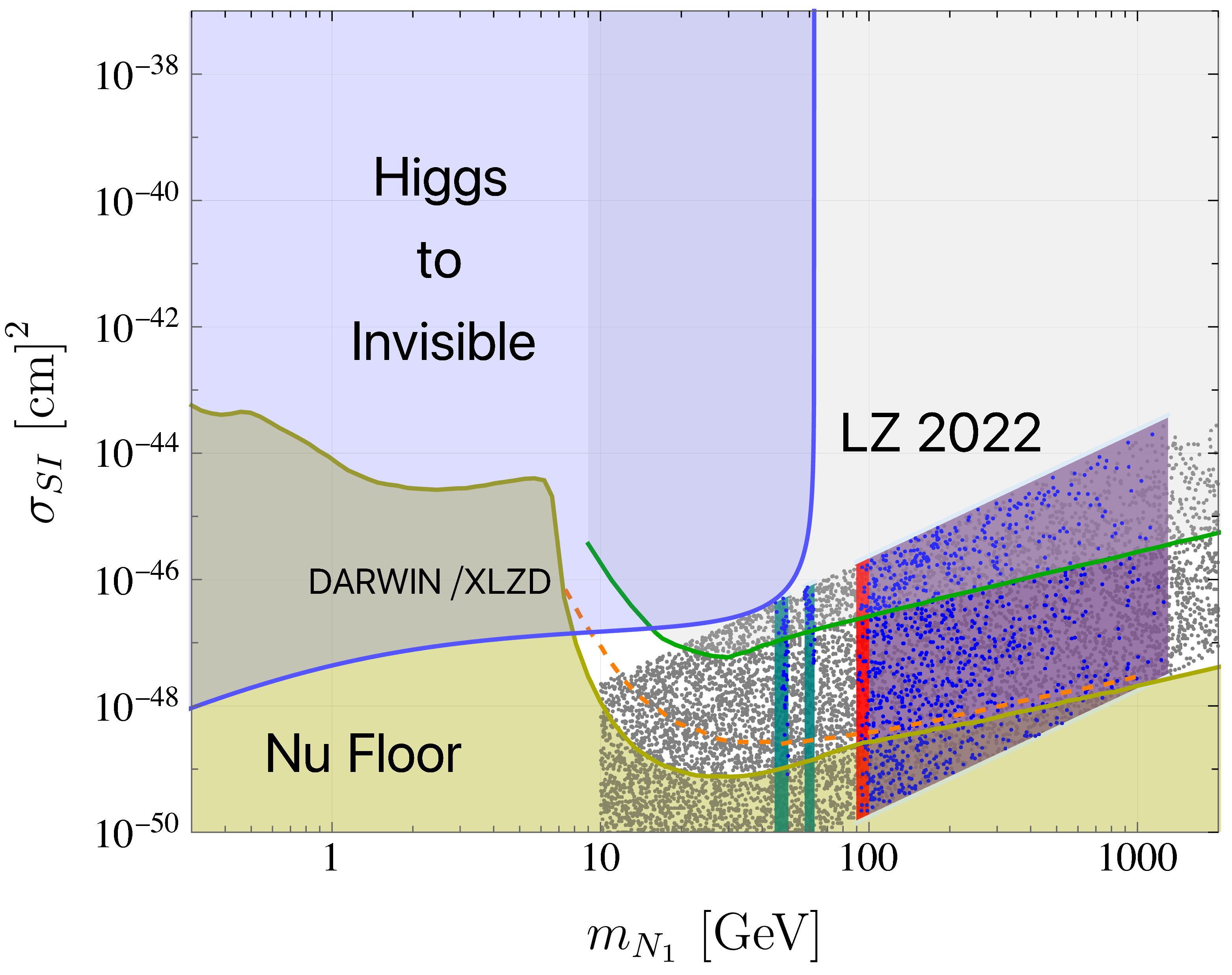}
        \caption{Direct detection in the model for $m_\varphi = 100 \, \rm GeV$. All data from Fig.~\ref{rd_100} are shown. Same color code as Fig.~\ref{P30}. In this case, the upper line of the bands is for $\lambda_5=10^{-2}$, and the lower line is for $\lambda_5=10^{-4}$. }
        \label{P100}
    \end{figure}

\begin{figure}[H]
        \centering
        \includegraphics[height=4in]{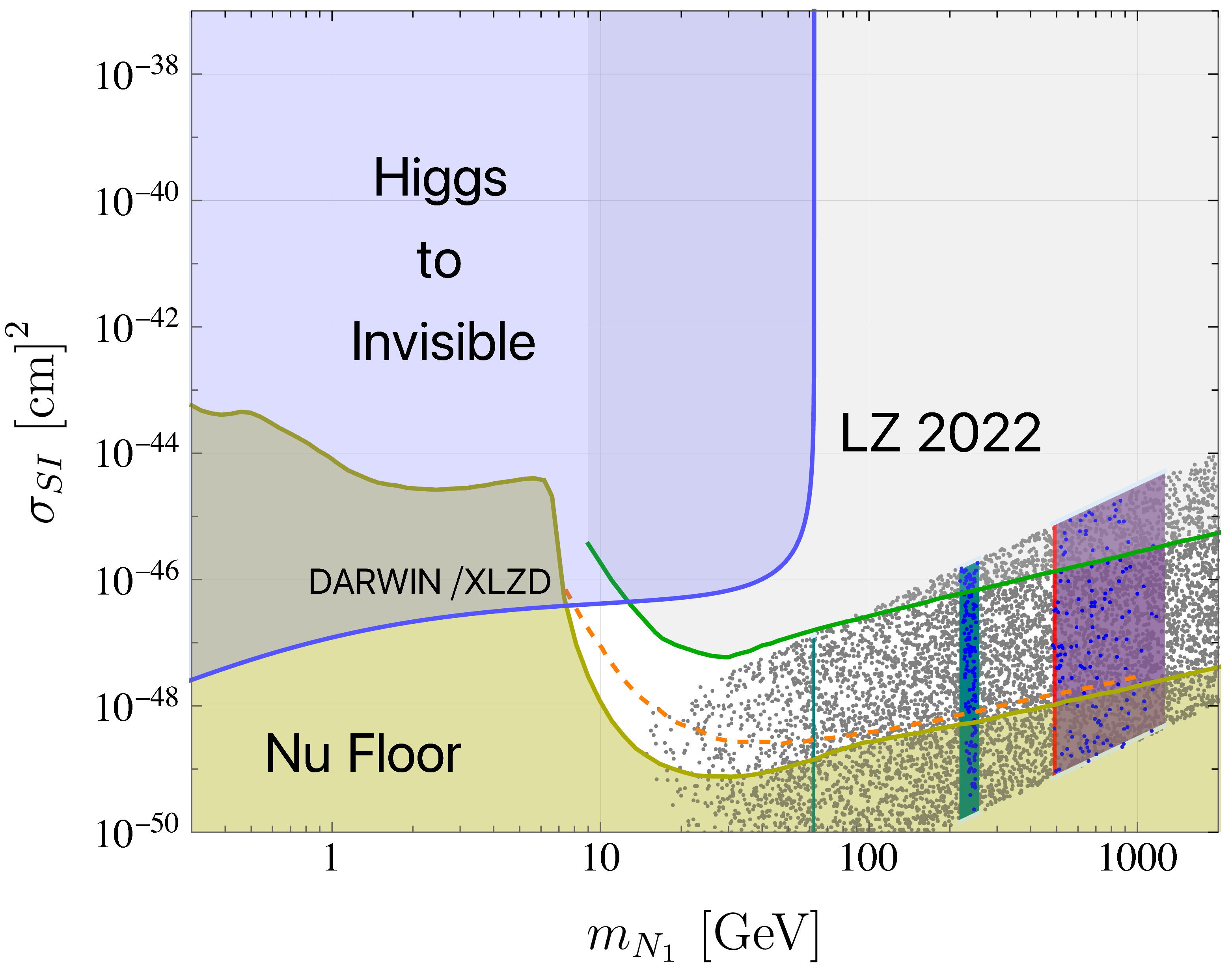}
        \caption{Direct detection in the model for $m_\varphi = 500 \, \rm GeV$. All data from Fig.~\ref{rd_500} are shown. Same color code as Fig.~\ref{P30}. In this case, the upper line of the bands is for $\lambda_5=10^{-1}$, and the lower line is for $\lambda_5=10^{-3}$. }
        \label{P500}
    \end{figure}

Our results for the experimental viability of data that produce both the observed DM relic abundance and generate experimentally consistent neutrino masses are shown in Figs.~\ref{P30}$-$\ref{P500}. Data are shown projected against the strongest current direct detection constraints from the LZ experiment \cite{LZ:2022lsv}. The currently strongest experimental constraint on Higgs-to-invisible decays\cite{ATLAS:2020kdi} is also shown in Figs.~\ref{P30}$-$\ref{P500}, although it is far less sensitive than direct detection limits at this DM mass scale.

The parameter region of our model, which produces the relic abundance of DM to within $3\sigma$ of the observed value, is indicated by colored regions in Figs.~\ref{P30}$-$\ref{P500}, and each distinct DM freeze-out regime is indicated by a distinct color. The purple region indicates where the relic abundance of DM is produced by kinematically allowed $NN \rightarrow \varphi \varphi$ annihilations, the red region indicates where the relic abundance is produced through kinematically suppressed $NN \rightarrow \varphi \varphi$ annihilations, and turquoise regions correspond to production of the DM abundance by $s$-channel DM-to-SM annihilations through an on-shell $h$ or $\varphi$ resonance.

The simultaneous generation of neutrino masses and thermal production of DM gives a wide and predictive experimentally viable window in the GeV$-$TeV mass range, below the sensitivity of current direct detection limits from LZ. Furthermore, the remaining experimental window above the neutrino floor of xenon will be nearly entirely explored by the near-future generation of direct detection experiments. Concretely, we show the projected sensitivity of DARWIN/XLZD\cite{Baudis:2024jnk}.

In particular, for the $m_\varphi=30, 100$ GeV cases we observe that the points with viable DM relic density produced through scalar resonances are excluded by LZ results or within reach of DARWIN. This result is not surprising, given that the annihilation cross section in these cases grows with $\sin \theta_V$, as is the case for direct detection. For the $m_\varphi=500$ GeV case, points with viable relic density are found beyond the reach of DARWIN and beneath the neutrino floor.

\section{Conclusions}
\label{sec:Conclusions}
We have presented a neutrino mass model with a DM candidate in the spirit of the scotogenic model. The purpose of this model is to show the viability of a simple one-loop neutrino mass model with fermion DM, where LFV constraints are decoupled from DM phenomenology, and no Majoron is introduced. To this end, we introduce a $\mathbb{Z}_4$ symmetry that is broken spontaneously to $\mathbb{Z}_2$. Since no global symmetry, imposed or accidental, is broken, no Goldstone boson appears in the spectrum. The dark sector fields generate neutrino masses at the one-loop level after electroweak and $\mathbb{Z}_{4}$ symmetry breaking. By setting a minimum value of the dark neutral scalar-pseudoscalar mass splitting, we can obtain the observed neutrino mass mixing and viable neutrino spectrum with small neutrino Yukawa couplings $Y_p<10^{-2}$, which guarantees unobservable LFV rates. The field breaking the discrete symmetry also introduces a channel for DM annihilation in the early Universe, which we have studied. We found that the new Yukawa coupling can lead to viable DM relic density in the kinematically allowed/forbidden regimes of $N N \rightarrow \varphi \varphi$ annihilations or in the scalar resonance regime $N N\xrightarrow{h,\phi} SM SM $. We have studied the prospects for DM detection in direct detection experiments, namely LZ and DARWIN. We observe that current LZ constraints have excluded part of the parameter space we consider, while DARWIN will continue probing the model. Some parts of the parameter space will remain untested after DARWIN and lie beneath the neutrino floor.\\
This allows us to conclude that scotogenic-type models of neutrino masses with light DM need not be in tension with current or future LFV experiments or cosmological constraints on light relics while having a thermal fermion DM candidate in the GeV$-$TeV scale.
$ \\ $
$ \\ $
\textit{Note Added.} Recently, a preprint appeared, presenting the same model but focusing on the scalar dark matter phenomenology \cite{Kim:2024cwp}. Additionally, their analysis explains the smallness of the effective $(\eta^\dagger H)^2$ coupling by integrating out the $\sigma$ field by taking $m_S>10^9 $ GeV. We have kept all scalar masses below $10$ TeV.

\section*{ACKNOWLEDGEMENTS}
This work is supported by the DGAPA UNAM Grant No. PAPIIT IN111625. The work of R.M.R. was supported by the EU Horizon 2020 research and innovation program, STRONG-2020 project, under Grant Agreement No. 824093, and by the Deutsche Forschungsgemeinschaft (DFG, German Research Foundation) – Project No. 514321794 (CRC1660: Hadron and Nuclei as discovery tools). The work of L.M.G.D.L.V. was supported by the National Science and Technology Council, the Ministry of Education (Higher Education Sprout Project No. NTU-113L104022-1), and the National Center for Theoretical Sciences of Taiwan.
R.M.R. acknowledges Consejo Nacional de Humanidades, Ciencias y Tecnologías (CONAHCYT) for funding. E.P. is grateful for the support of Programa de Apoyos para la Superación del Personal Académico (PASPA)-DGAPA, UNAM for a sabbatical leave, and Fundaci\'on Marcos Moshinsky. \\

\bibliography{references}

\end{document}